\documentclass[onecolumn,floatfix,superscriptaddress,showpacs,showkeys,nofootinbib,preprint]{revtex4}
\usepackage{epsfig}
\usepackage{amssymb,latexsym,amsmath}
\newcommand{\eq}[1]{\begin{align} #1 \end{align}}

\begin{document}

\title{Strongly Intensive Measures for Multiplicity Fluctuations}

\author{V.V. Begun}
 \affiliation{Bogolyubov Institute for Theoretical Physics, Kiev, Ukraine}
 \affiliation{Frankfurt Institute for Advanced Studies, Frankfurt,Germany}

 \author{V.P. Konchakovski}
 \affiliation{Bogolyubov Institute for Theoretical Physics, Kiev, Ukraine}
 \affiliation{Institute for Theoretical Physics, University of Giessen, Germany}

 \author{M.I. Gorenstein}
 \affiliation{Bogolyubov Institute for Theoretical Physics, Kiev, Ukraine}
 \affiliation{Frankfurt Institute for Advanced Studies, Frankfurt,Germany}

 \author{E.L. Bratkovskaya}
 \affiliation{Frankfurt Institute for Advanced Studies, Frankfurt, Germany}
 \affiliation{Institut for Theoretical Physics, University of Frankfurt, Frankfurt, Germany}

%
\begin{abstract}

The recently proposed two families of strongly intensive measures
of fluctuations and correlations are studied within
Hadron-String-Dynamics (HSD) transport approach to nucleus-nucleus
collisions. We consider the measures $\Delta^{K\pi}$ and
$\Sigma^{K\pi}$ for kaon and pion multiplicities in Au+Au
collisions in a wide range of collision energies and centralities.
These strongly intensive measures appear to cancel the participant
number fluctuations. This allows to enlarge the centrality window
in the analysis of event-by-event fluctuations up to at least of
10\% most central collisions. We also present a comparison of the
HSD results with the data of NA49 and STAR collaborations. The HSD
describes $\Sigma^{K\pi}$ reasonably well. However, the HSD
results depend monotonously on collision energy and do not
reproduce the bump-deep structure of $\Delta^{K\pi}$ observed from
the NA49 data in the region of the center of mass energy of
nucleon pair $\sqrt{s_{NN}}= 8\div 12$~GeV. This fact deserves
further studies. The origin of this `structure' is not connected
with simple geometrical or limited acceptance effects, as these
effects are taken into account in the HSD simulations.

\end{abstract}

\pacs{12.40.-y, 12.40.Ee}

\keywords{event-by-event fluctuations, nucleus-nucleus collisions}

\maketitle

\section{Introduction}
A possibility to observe signatures of the critical point of QCD
matter inspired the energy and system size scan programs of the
NA61 collaboration at the SPS CERN~\cite{Gazdzicki:NA61} and the
low energy scan program of STAR and PHENIX collaborations at the
RHIC BNL~\cite{RHIC-SCAN}. These experimental studies focus on the
event-by-event (e-by-e) fluctuation measurements in
nucleus-nucleus (A+A) collisions. One should compare the
fluctuation properties of hadrons produced in  collisions of
different nuclei at different collision energies. In these
reactions the average sizes of the created physical systems and
their e-by-e fluctuations are rather different~\cite{KGBG:2010}.
The fluctuations of the system volume strongly affect the observed
hadron fluctuations, i.e. the measured hadron fluctuations do not
describe the physical properties of the system but rather reflect
the system size fluctuations. In A+A collisions with different
centralities a system volume is indeed changed significantly from
interaction to interaction. These event-by-event volume variations
of the produced matter are usually out of the experimental
control.

We recall that extensive quantities are proportional to the system
volume $V$, whereas  intensive quantities  do not. They are used
to describe the local properties of a physical system. In
particular, an equation of state of the matter is usually
formulated in terms of the intensive physical quantities, e.g.,
the pressure is considered as a function of temperature and
chemical potentials.

In statistical physics a mean value $\langle N\rangle$ of a
fluctuating number of particles is an  extensive quantity, i.e.,
$\langle N\rangle\propto V$, whereas the ratio of mean
multiplicities of two different particle types is an intensive
quantity. If local properties of the system remain
unchanged\footnote{ This is approximately valid in a wide range of
centralities, and violated only for  very peripheral collision
events~\cite{STAR-T-mu}.}, this ratio does not depend on the
average size of the system and of its fluctuations.
Particle number fluctuations are quantified by the variance,
Var$(N)=\langle N^2 \rangle - \langle N \rangle^2$, which is an
extensive quantity  in statistical models, while the scaled
variance, \mbox{$\omega_N=[\langle N^2 \rangle - \langle N
\rangle^2] / \langle N \rangle$}, is an intensive one. However,
the scaled variance being an intensive quantity  depends on the
system size fluctuations.

In the event-by-event analysis of A+A collisions the number of
nucleon participants $N_{part}$ and the scaled variance of its
fluctuations $\omega_{part}$ play the same role as the volume and
volume fluctuations in statistical models. For example, the
fluctuations of nucleon participants strongly contribute to the
scaled variances of  charged particles~\cite{WA98}, pions and
kaons~\cite{KGBG:2010}. To avoid these unnecessary contributions
one needs to make a very rigid centrality selection. The analysis
of the scaled variances has to be limited to about 1\% most
central events only (see, e.g., Ref.~\cite{Lung}). This causes two
problems. First, there are technical problems with a strict
centrality selection. Second, for a more rigid centrality trigger
one evidently loses the number of collision events and thus needs
to enlarge strongly the total event statistics.

%
The analysis of fluctuations of hadron production
properties in collisions of relativistic nuclei  may profit
from the use of measurable intensive quantities which are
independent of both the average size of the system  and of the
size variations. Two families of these quantities -- referred to
as strongly intensive ones -- have been recently proposed in
Ref.~\cite{GG:2011}.
In the present study we consider the properties of these strongly
intensive measures for particle number fluctuations in A+A
collisions within the Hadron-String-Dynamics (HSD) transport
approach~\cite{HSD}. We use HSD as it describes well the particle
spectra in heavy ion experiments. The large fluctuations of
nucleon participants number is under theoretical control. Within
the HSD simulations we can estimate and separate these unnecessary
fluctuations. Besides, we can check whether these system size
fluctuations are really cancelled out in strongly intensive
measures.

The strongly intensive measures $\Delta^{AB}$ and
$\Sigma^{AB}$~\cite{GG:2011} can be defined for two arbitrary
extensive quantities $A$ and $B$. To be specific we consider the
total hadron multiplicities of charged kaons $K=K^++K^-$  and
pions\footnote{Another pair of extensive quantities -- particle
multiplicity and sum of their transverse momenta modules -- within
the UrQMD transport model have been recently discussed in
Ref.~\cite{Phi-pT}} $\pi=\pi^++\pi^-$
:
 \eq{\label{Delta}
 \Delta^{K\pi}
 ~&=~ \frac{1}{\langle K\rangle+\langle \pi\rangle}~ \big[ ~\langle \pi\rangle~
      \omega_K ~-~\langle K\rangle ~\omega_\pi ~\big]~,
 \\
  \Sigma^{K\pi}
 ~&=~ \frac{1}{\langle K\rangle+\langle \pi\rangle}~\big[~
      \langle \pi\rangle~\omega_K ~+~\langle K\rangle~ \omega_\pi ~-~2\left(
      \langle K\pi \rangle -\langle K\rangle\langle
      \pi\rangle\right)~\big]~,
 \label{Sigma}
 }
where
 \eq{\label{Omega}
 \omega_K ~\equiv~ \frac{\langle K^2\rangle ~-~ \langle K\rangle^2}{\langle
 K\rangle}~,~~~~
 \omega_\pi ~=~ \frac{\langle \pi^2\rangle ~-~ \langle \pi\rangle^2}{\langle
 \pi\rangle}~
 }
are the scaled variances of the $K$ and $\pi$ fluctuations.

The paper is organized as follows. In Section \ref{sec-MIS} we
consider the properties of different fluctuation measures within
the model of independent sources. Section \ref{sec-HSD} presents
the HSD results in A+A collisions. In Section \ref{sec-Exp} a
comparison of the HSD results with the available data is
presented. A summary in Section \ref{sec-sum} closes the article.

\section{Model of Independent Sources}\label{sec-MIS}
It is instructive to start from the model of independent sources
(MIS) for multi-particle production in A+A collisions. The number
of sources in this model changes from event to event. However, the
sources are assumed to be statistically identical (i.e., the
average properties of all sources are the same) and independent
(i.e., there are no correlations between hadrons produced from
different sources).
The first and the most popular example of the model of independent
sources is the Wounded Nucleon Model~\cite{WNM}. In this model,
one assumes that A+A collision can be treated as a superposition
of independent contributions from each of $N_{part}$ nucleon
participants. For example, the kaon and pion multiplicities are
the following:
\eq{\label{MIS}
K~=~K_1+K_2+\ldots+K_{N_{part}}~,~~~~~\pi~=~\pi_1+\pi_2+\ldots+\pi_{N_{part}}~.
}
The average multiplicities from each source are equal:
\eq{\label{sources}
\langle K_1\rangle=\langle K_2\rangle=\ldots=\langle
K_{N_{part}}\rangle~\equiv~n_K~,~~~~ \langle \pi_1\rangle=\langle
\pi_2\rangle=\ldots=\langle \pi_{N_{part}}\rangle~\equiv~n_\pi~.
}
Thus, the e-by-e averages of final hadron multiplicities can be
obtained as:
\eq{\label{mult-K-MIS}
&\langle
K\rangle~=~\sum_{N_{part}}\,P(N_{part})\,\sum_{j=1}^{N_{part}}\,\langle
K_j\rangle ~=~\sum_{N_{part}}\,P(N_{part})\,n_K\,N_{part}~ =~ n_K
\cdot \langle N_{part}\rangle~,\\
& \langle
\pi\rangle~=~\sum_{N_{part}}\,P(N_{part})\,\sum_{j=1}^{N_{part}}\,\langle
\pi_j\rangle ~=~\sum_{N_{part}}\,P(N_{part})\,n_\pi\,N_{part}~ =~
n_\pi \cdot \langle N_{part}\rangle~,\label{mult-pi-MIS}
}
where $\langle N_{part}\rangle$ is the average number of nucleon
participants (i.e., wounded nucleons). The quantities $n_K$ and
$n_\pi$ in Eqs.~(\ref{mult-K-MIS},\ref{mult-pi-MIS}) are the
average multiplicities per one nucleon participant.
For the second moment of
kaon multiplicity distributions one obtains:
 \eq{\label{K1K2}
 \langle K^2 \rangle  = \sum_{N_{part}}\,P(N_{part})\,
 \langle\,\left(\sum_{i=1}^{N_{part}} K_i\right)^2 \,\rangle
 = \sum_{N_{part}}\,P(N_{part})
 \left[ \sum_{i=1}^{N_{part}} \langle K_i^2 \rangle  + \sum_{1\neq i<j\leq
 N_{part}}
 \langle K_iK_j \rangle\,\right] .
 }
The right hand side of Eq.~(\ref{K1K2}) is a sum of the
$N_{part}^2$ terms $\langle K_iK_j\rangle$. The number of terms
with $i=j$ is $N_{part}$,  and the number of ones with $i\neq j$
is $N_{part}^2-N_{part}$. The different sources are assumed to be
statistically identical, i.e. the second moments of kaon number
distributions from the different sources are equal to each other:
 \eq{\label{Ki2}
 \langle K^2_i\rangle
 ~=~\langle K^2_1\rangle~,
 }
with $i=2,\ldots,N_{part}$. The different  sources are also
assumed to be independent, i.e. the kaon-pion pairs emitted by
different sources are uncorrelated. This gives for $i\neq j$:
\eq{\label{KiKj}
\langle K_iK_j\rangle~=~\langle K_i\rangle \langle K_j\rangle~=~ n_K^2~.
 }
From the above equations one finds:
\eq{\label{MIS-K2}
\langle K^2\rangle ~=~\langle K^2_1\rangle \,\langle
N_{part}\rangle ~+~ n_K^2\,\Big[\,\langle
N_{part}^2\rangle~-~\langle N_{part}\rangle\,\Big]~.
}
Similarly, one obtains for pions:
 \eq{
\langle \pi^2\rangle ~&=~\langle \pi^2_1\rangle\,\langle
N_{part}\rangle ~+~ n_\pi^2\,\Big[\,\langle
N_{part}^2\rangle~-~\langle
N_{part}\rangle\,\Big]~.\label{MIS-pi2}
 }
For the kaon-pion correlations one finds
\eq{\label{K-pi-i}
\langle K_i\pi_i \rangle ~=~\langle K_1\pi_1 \rangle
%
}
for $i=2,\ldots,N_{part}$, and
\eq{\label{K-pi-dif}
\langle K_i\pi_j\rangle~=~n_K\,n_{\pi}
}
for $i\neq j$.  It then follows:
 \eq{
 \langle K\pi\rangle ~&=~\langle K_1\pi_1\rangle\,\langle
N_{part}\rangle ~+~ n_K\,n_\pi\,\Big[\,\langle
N_{part}^2\rangle~-~\langle
N_{part}\rangle\,\Big]~.\label{MIS-Kpi}
}
The scaled variances for the production of kaons and pions in MIS
are then presented as:
%
 \eq{\label{omegaWNM}
 \omega_K ~=~ \omega_K^* ~+~ n_K~\omega_{part}~,~~~~~~
 \omega_\pi ~=~ \omega_\pi^* ~+~ n_\pi~\omega_{part}~,
 }
where $\omega_K^*$ and $\omega_\pi^*$ are, respectively, the scaled
variances of kaons and pions from one source,
\eq{\label{omega*}
\omega_K^*~=~\frac{\langle K^2_1\rangle~-~\langle K_1\rangle^2}
{\langle K_1\rangle}~,~~~~
\omega_\pi^*~=~\frac{\langle \pi^2_1\rangle~-~\langle
\pi_1\rangle^2} {\langle \pi_1\rangle}~,
}
and $\omega_{part}$ is the scaled variance of e-by-e fluctuations
of the number of nucleon participants,
\eq{\label{omegas}
\omega_{part}~=~\frac{\langle N_{part}^2\rangle~-~\langle
N_{part}\rangle^2}{\langle N_{part}\rangle}~.
}
Similar to Eq.~(\ref{omegaWNM}) one finds within MIS,
\eq{\label{rho-MIS}
%
\frac{\langle K\,\pi\rangle~-~\langle K\rangle\,\langle
\pi\rangle}{\langle K~+~\pi\rangle} ~=~
\frac{\rho^*_{K\pi}}{n_K+n_\pi}~+~ \frac{n_K\,n_\pi}{n_K+n_\pi}~
\omega_{part}~,
}
where
\eq{\label{rho*}
\rho^*_{K\pi}~\equiv~\langle K_1\pi_1\rangle~-~\langle
K_1\rangle\,\langle \pi_1\rangle
}
describes the correlations between $K$ and $\pi$ numbers  in one
source.

In MIS, the scaled variances $\omega_K$ and $\omega_\pi$
(\ref{omegaWNM}) are independent of the average number of nucleon
participants $\langle N_{part}\rangle$. Thus, $\omega_K$ and
$\omega_\pi$ are intensive quantities. However, they depend on the
fluctuations of the number of nucleon participants via
$\omega_{part}$ and, therefore, they are not strongly intensive
quantities.
From above formulas it follows that both measures $\Delta$
(\ref{Delta}) and $\Sigma$ (\ref{Sigma}) are strongly intensive
quantities within the MIS, i.e. they are independent of $\langle
N_{part}\rangle $ and of $\omega_{part}$~:
\eq{\label{Delta1}
\Delta^{K\pi}~&=~\frac{1}{n_K+n_\pi}
~\left[~n_\pi~\omega_K^*~-~n_K~\omega_\pi^*~\right]~,\\
\Sigma^{K\pi}~&=~\frac{1}{n_K+n_\pi} ~\left[~n_\pi~\omega^*_K
~+~n_K~\omega^*_\pi ~-~2\rho^*_{K\pi}~\right] ~.\label{Sigma1}
}

The contributions from nuclear participants can be obtained from
nucleon-nucleon collisions which should be considered as the
properly weighted sum of p+p, p+n, and n+n
interactions~\cite{KGB:2007}. At high SPS and RHIC energies the
results in nucleon-nucleon collisions are close to those in p+p
collisions. Thus, in the above equations one may approximate the
MIS quantities with the results of p+p inelastic interactions at
the same collision energy per nucleon. Inelastic p+p collisions
might be understood within MIS as the system with $N_{part}=2$ and
$\omega_{part}=0$. It then follows:
\eq{\label{n-pp}
 n_K & \cong \frac{1}{2}\,\langle K\rangle_{pp}~,~~~~~ n_\pi \cong
\frac{1}{2}\, \langle \pi\rangle_{pp}~,~~~~\rho^*_{K\pi}~\cong~
\frac{1}{2}\,\big[\langle K\,\pi\rangle_{pp}~-~ \langle
K\rangle_{pp} \,\langle \pi \rangle_{pp}\big]
~,\\
\omega_K^* & \cong \frac{\langle K^2\rangle_{pp}-\langle
K\rangle_{pp}^2}{\langle K\rangle_{pp}}~,~~~~~~\omega_\pi^*\cong
\frac{\langle \pi^2\rangle_{pp}-\langle \pi\rangle_{pp}^2}{\langle
\pi\rangle_{pp}}~,\label{omega-pp}
}
i.e., particle multiplicities $n_K$ and $n_\pi$, and
$K$-$\pi$-correlations $\rho^*_{K\pi}$ from one source are equal
to one half of those values in p+p collisions, whereas the scaled
variances $\omega_K^*$ and $\omega^*_\pi$ for one source coincide
with the corresponding scaled variances in p+p collisions.

Another interpretation of MIS can be obtained in terms of the
statistical mechanics. One may assume that the matter created in
A+A collisions at different centralities corresponds to systems in
statistical equilibrium with the same temperature and chemical
potentials, but with a volume varying from event to event. It is
also natural to assume that the volume is proportional to the
number of nucleon participants $V\propto N_{part}$. One then
finds~\cite{GG:2011} within the grand canonical ensemble
formulation a validity of Eqs.~(\ref{Delta1}) and (\ref{Sigma1}).
In this case, the hadron multiplicities $n_K$ and $n_\pi$, scaled
variances $\omega_K^*$ and $\omega_\pi^*$, and correlation
$\rho_{K\pi}^*$ should be found as the corresponding quantities at
fixed volume (i.e., at fixed $N_{part}$). The transport model
calculations demonstrate~\cite{KGBG:2010} that the e-by-e
fluctuations of the number of participants become negligible in
the sample of most central A+A collisions, e.g., one finds
$\omega_{part}\ll 1$ in Pb+Pb (or Au+Au) collisions with impact
parameter equal to zero, $b=0$. Therefore, one may define the
parameters of MIS as:
\eq{\label{n-b0}
& n_K~=~\frac{\langle K\rangle_{b=0}}{\langle
N_{part}\rangle_{b=0}}~,~~~~ n_\pi~=~\frac{\langle
\pi\rangle_{b=0}}{\langle
N_{part}\rangle_{b=0}}~,~~~~&\rho_{K\pi}^*~\cong~\frac{\langle
K\,\pi\rangle_{b=0}~-~\langle K\rangle_{b=0}~ \langle
\pi\rangle_{b=0}}{\langle N_{part}\rangle_{b=0}}~,\\
& \omega_{\pi}^*~\cong~\omega_{\pi}(b=0)~,~~~~~~ \omega_K^*~\cong~
\omega_K(b=0)~,\label{omega-b0}
}
i.e., particle multiplicities $n_K$ and $n_\pi$, and
$K$-$\pi$-correlations $\rho^*_{K\pi}$ are equal to the results in
A+A collisions with zero impact parameter divided by the average
number of nucleon participants at $b=0$.  On the other hand, the
scaled variances $\omega_K^*$ and $\omega^*_\pi$ entering in MIS
results just coincide with the corresponding scaled variances in
A+A collisions at $b=0$.

Another quantity frequently used to characterize the fluctuations
of $K$ and $\pi$ particle numbers is~\cite{Niu}:
\eq{\label{nu}
%
\nu^{K\pi}_{dyn}~\equiv~\frac{\langle K(K-1)\rangle}{\langle
K\rangle^2}~+~\frac{\langle \pi(\pi-1)\rangle}{\langle
\pi\rangle^2}~-~2~\frac{\langle K\pi\rangle}{\langle K\rangle\langle
\pi\rangle}~.
}
One can easily find the
relation:
\eq{\label{nu-Psi}
\nu^{K\pi}_{dyn}~=~\frac{\langle K+\pi\rangle}{\langle K \rangle
\langle \pi\rangle }\left[~\Sigma^{K\pi}~-~1\right]~.
%
}
Equation (\ref{nu-Psi}) shows that $\nu^{AB}_{dyn}$~, similar to
$\Sigma^{AB}$, is independent of fluctuations of the
number of participants, but it decreases as $\nu_{dyn}^{AB} \propto \langle
N_{part}\rangle ^{-1}$ and, thus, it is not an intensive quantity.
We prefer to use the strongly intensive quantities and will
compare the HSD results with the data on $\nu_{dyn}$ but
recalculated in $\Sigma$ according to Eq.~(\ref{nu-Psi}).

\section{The Results of Hadron-String-Dynamics}\label{sec-HSD}
\begin{figure}[ht!]
 \epsfig{file=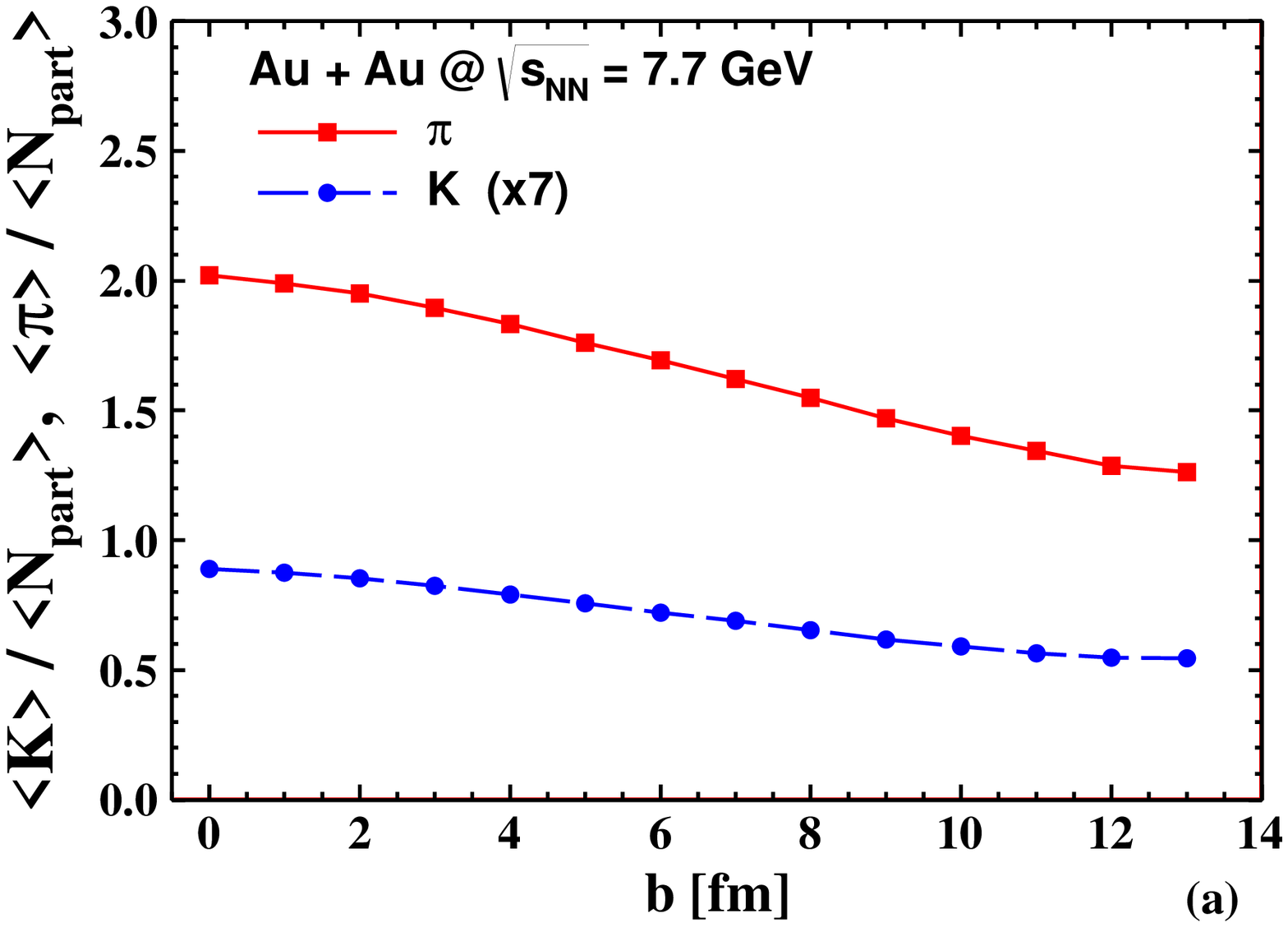,width=0.45\textwidth}
\epsfig{file=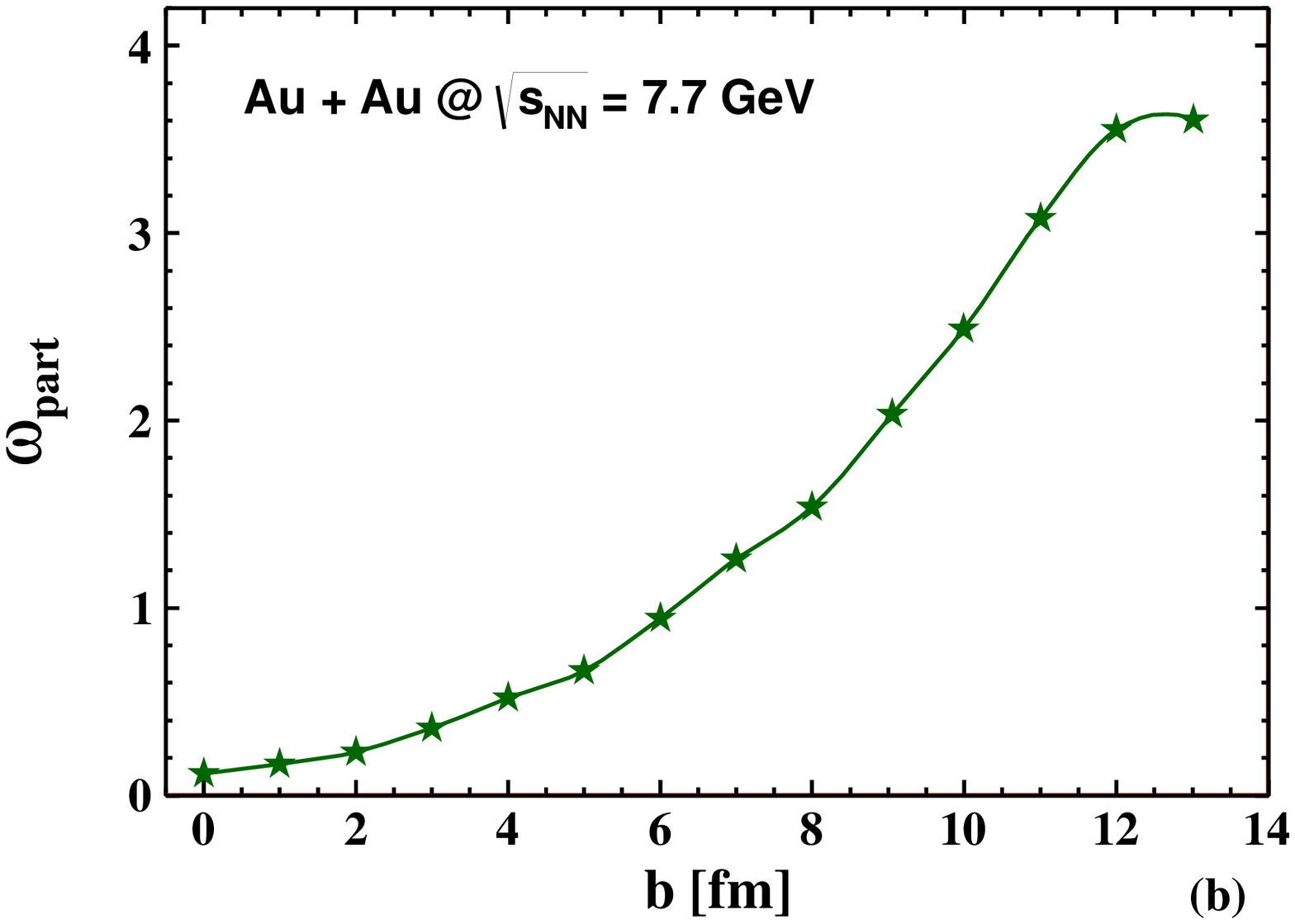,width=0.45\textwidth}
\epsfig{file=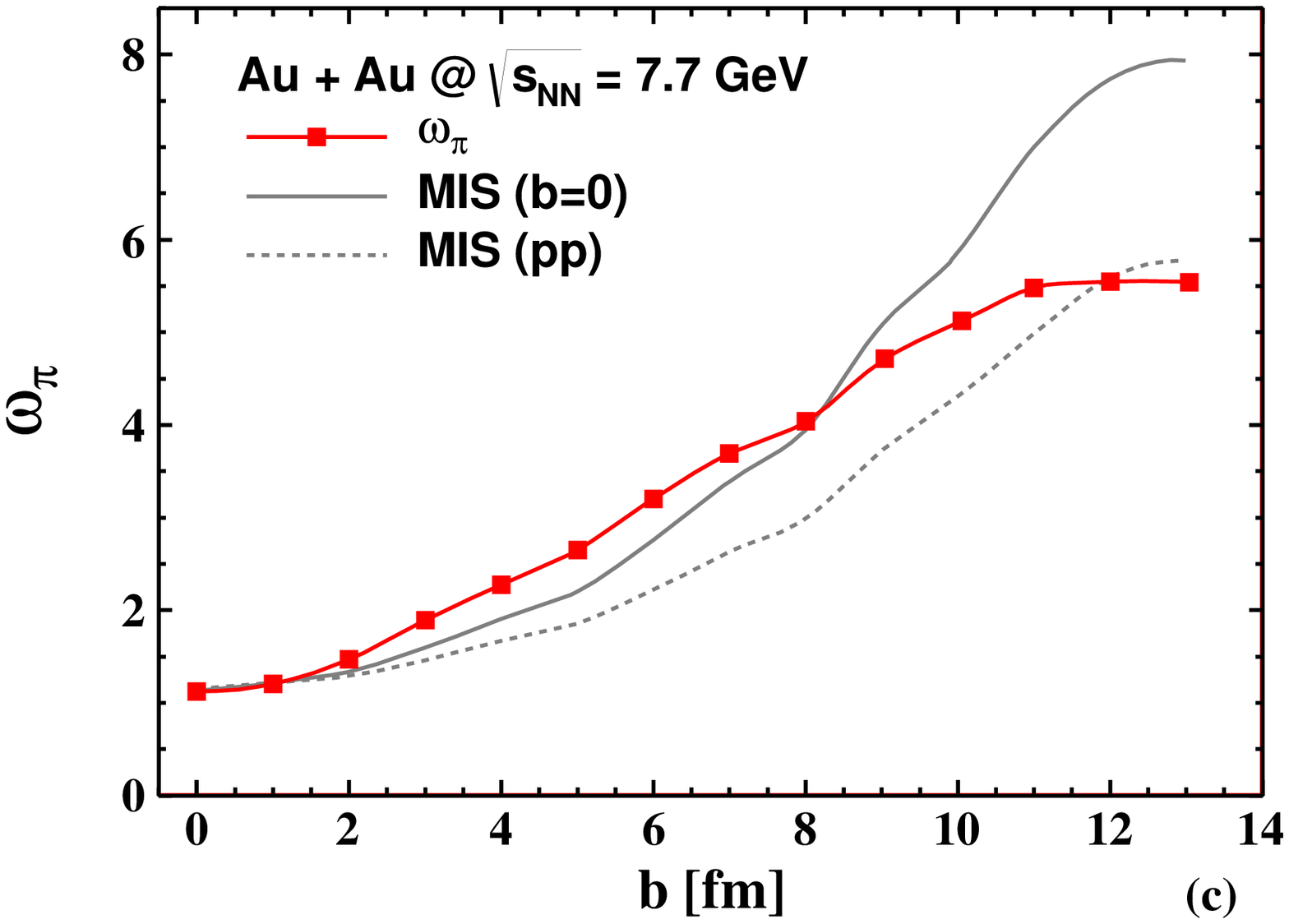,width=0.45\textwidth}
\epsfig{file=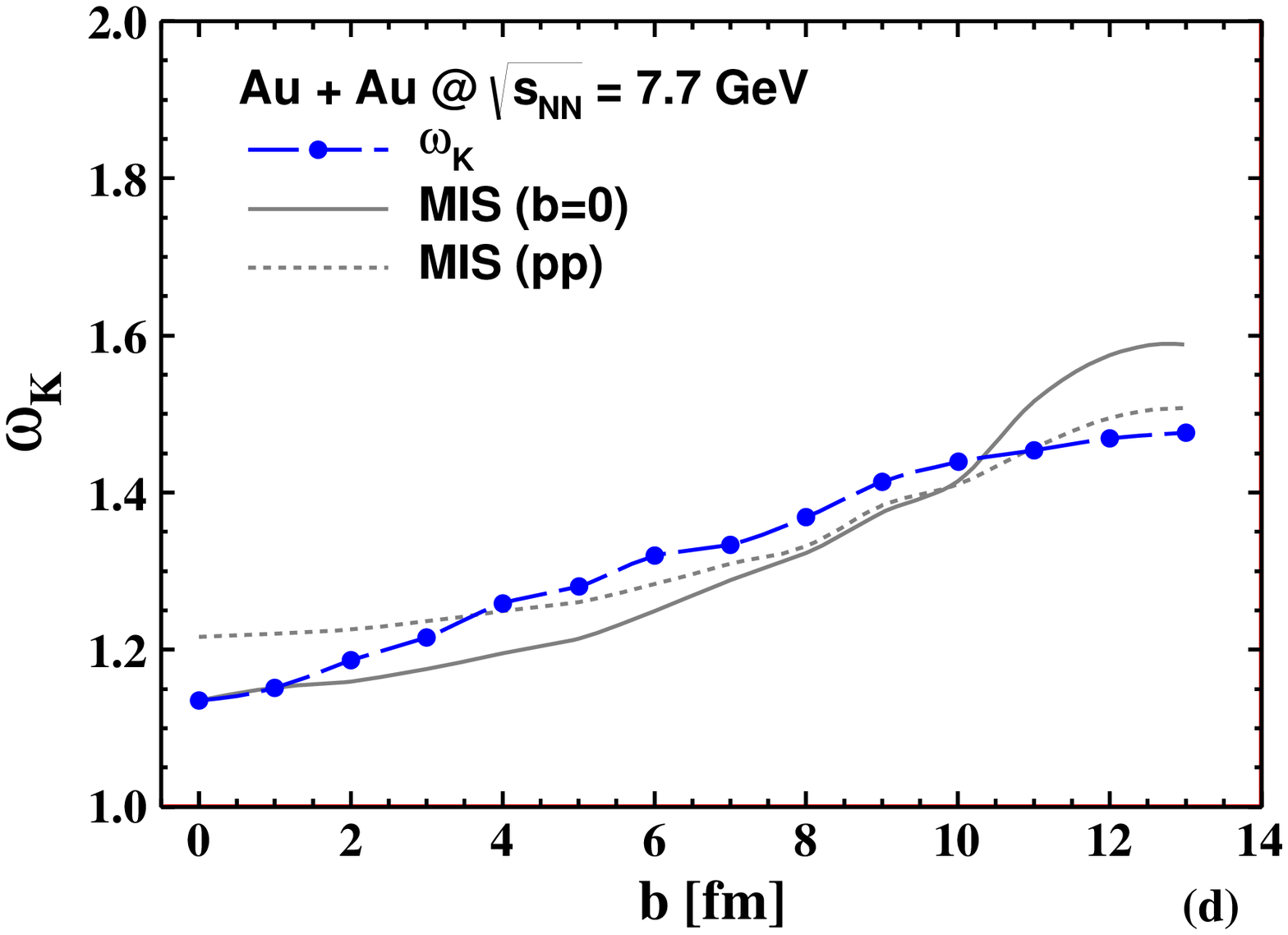,width=0.45\textwidth}
 \caption{
The symbols correspond to the HSD results at different impact
parameter $b$ in Au+Au collisions at $\sqrt{s_{NN}}=7.7$~GeV. (a):
The HSD ratio of pion and kaon multiplicities to the average
number of participants. Note that $\langle K\rangle/\langle
N_{part}\rangle$ is multiplied by a factor of 7. (b): The scaled
variance $\omega_{part}$. (c): The scaled variance $\omega_\pi$.
(d): The scaled variance $\omega_K$. The solid and dotted lines in
(c) and (d) show the MIS($b=0$) and MIS(pp) results, respectively.
} \label{fig1}
\end{figure}

In order to study the properties of  strongly intensive measures
(\ref{Delta}) and (\ref{Sigma}) we calculate the fluctuations of
kaon and pion numbers with the HSD transport approach~\cite{HSD}.
First, we consider the centrality dependence in A+A collisions.
The HSD results in Au+Au collisions at the center of mass energy
of the nucleon pair $\sqrt{s}_{NN}=7.7$~GeV ($E_{lab}\simeq 30$
AGeV) are presented in Fig.~\ref{fig1}.

\begin{figure}[ht!]
 \epsfig{file=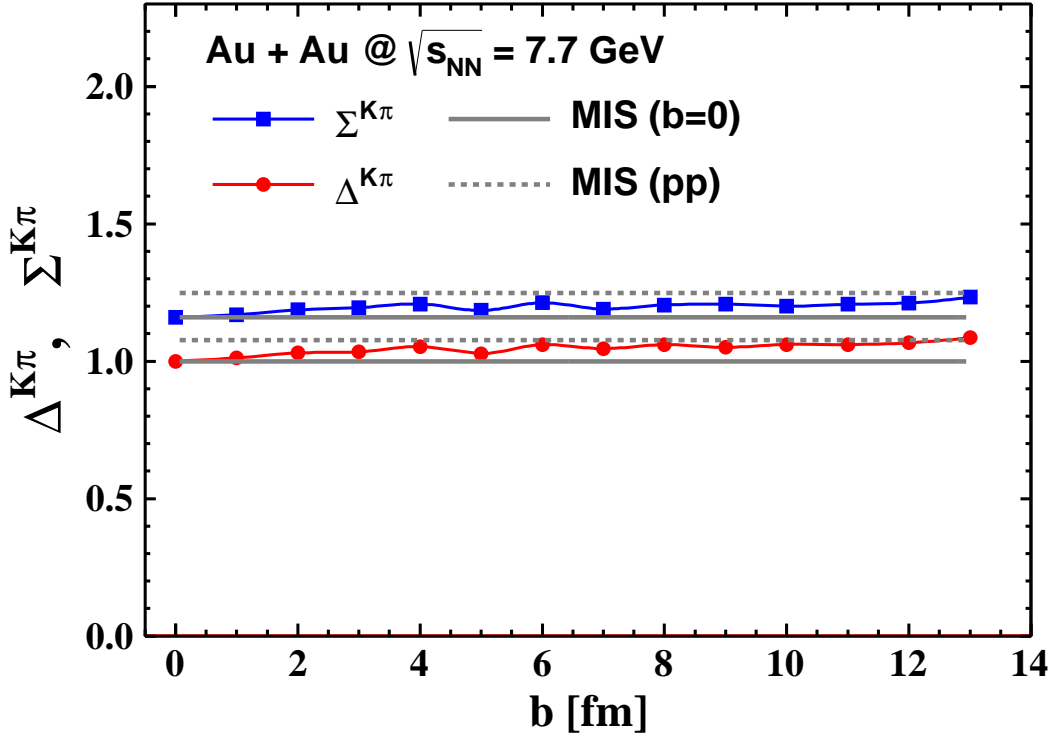,width=0.9\textwidth}
 \caption{
 The strongly intensive measures
 $\Delta^{K\pi}$ 
 (circles) and $\Sigma^{K\pi}$ 
 (squares). The symbols correspond to the HSD results
for Au+Au collisions at $\sqrt{s_{NN}}=7.7$~GeV. The horizontal
solid lines show the MIS($b=0$) results for $\Delta^{K\pi}$ and
$\Sigma^{K\pi}$. The horizontal dotted lines show the MIS(pp)
results. The lower lines correspond to $\Delta^{K\pi}$ and the
upper lines to $\Sigma^{K\pi}$. }\label{fig2}
\end{figure}

\begin{figure}[ht!]
 \epsfig{file=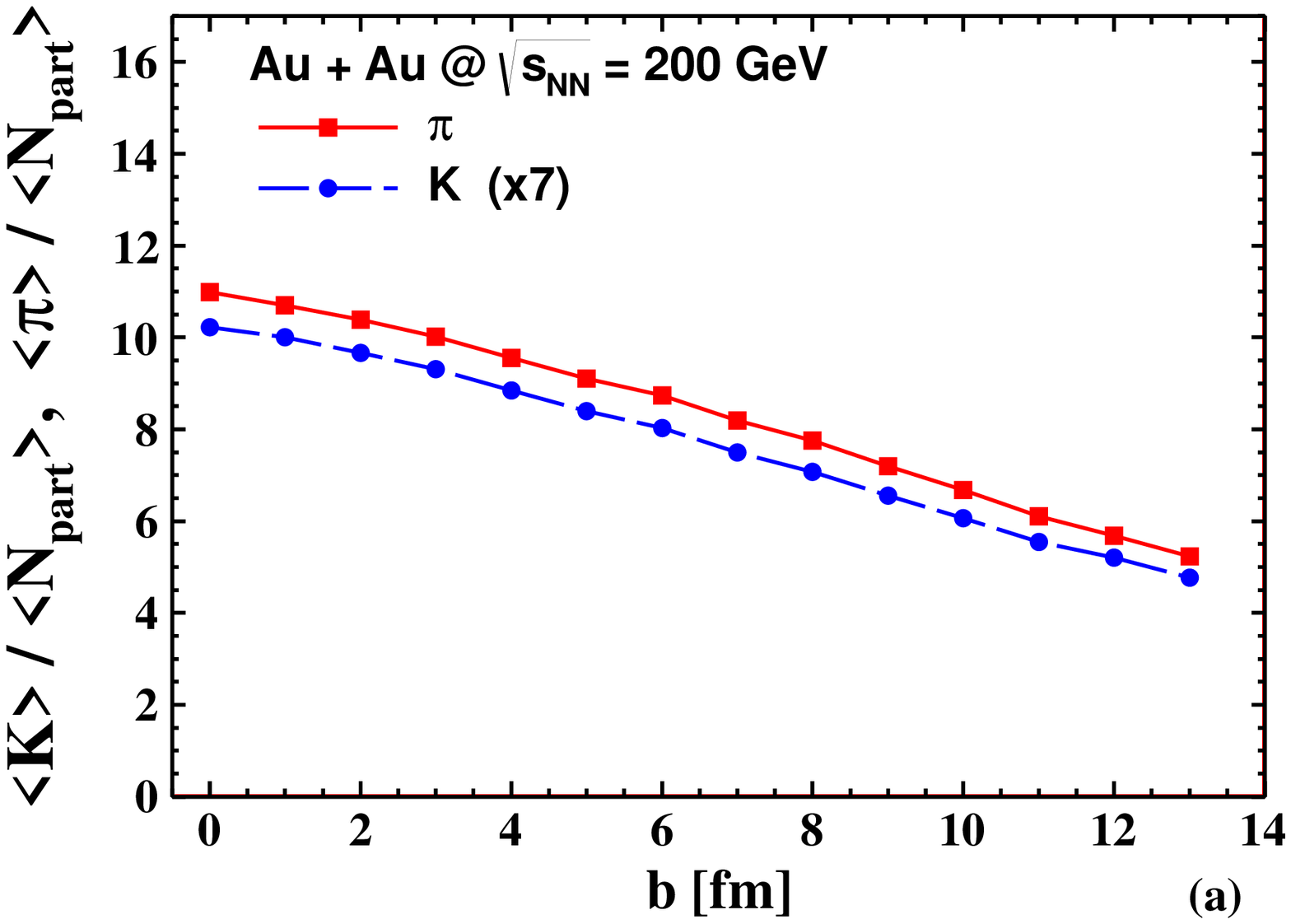,width=0.45\textwidth}
 \epsfig{file=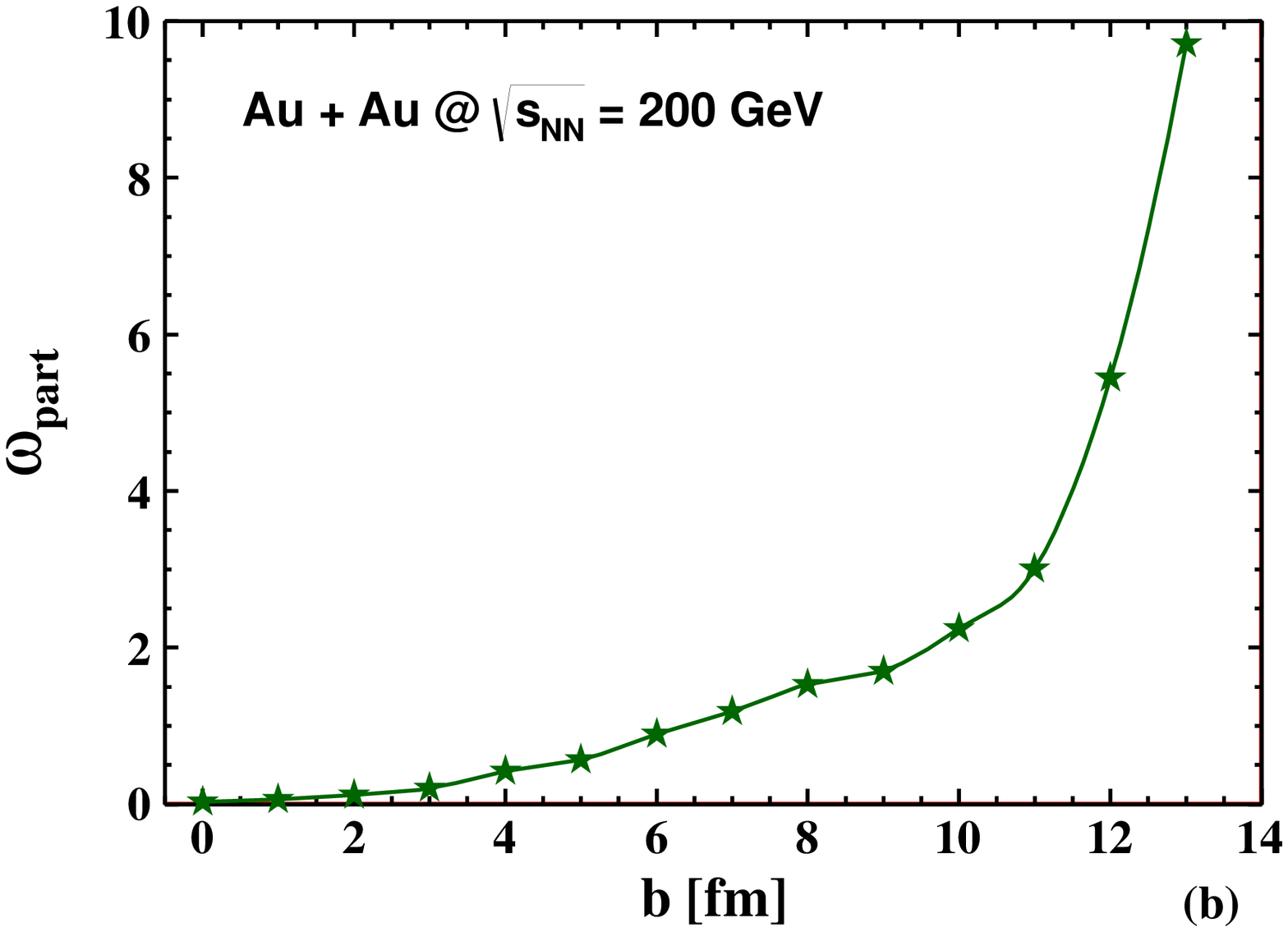,width=0.45\textwidth}
 \epsfig{file=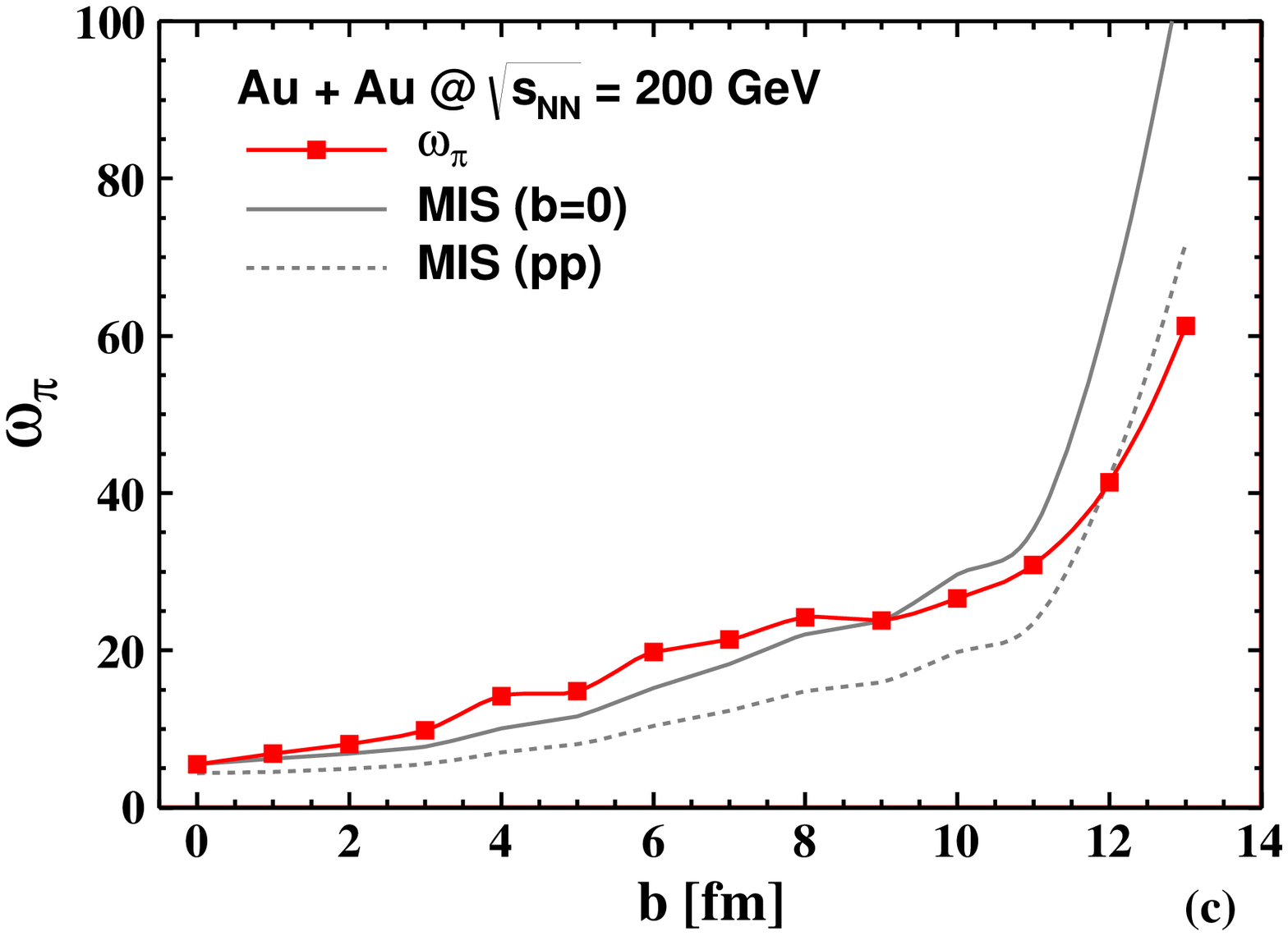,width=0.45\textwidth}
 \epsfig{file=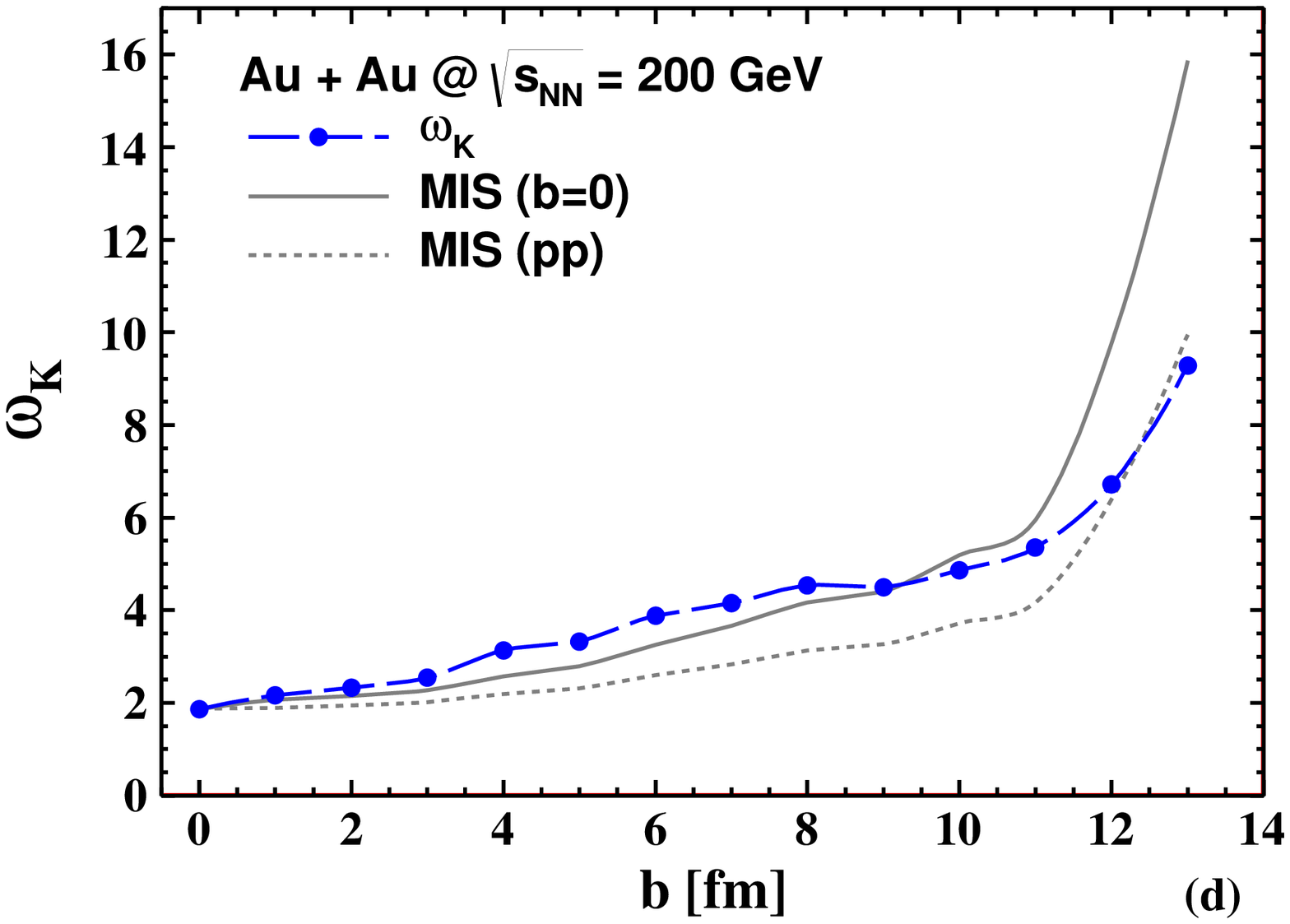,width=0.45\textwidth}
 \caption{
 The same as in Fig.~\ref{fig1}
but for $\sqrt{s_{NN}}=200$~GeV.}\label{fig3}
\end{figure}

\begin{figure}[ht!]
\epsfig{file=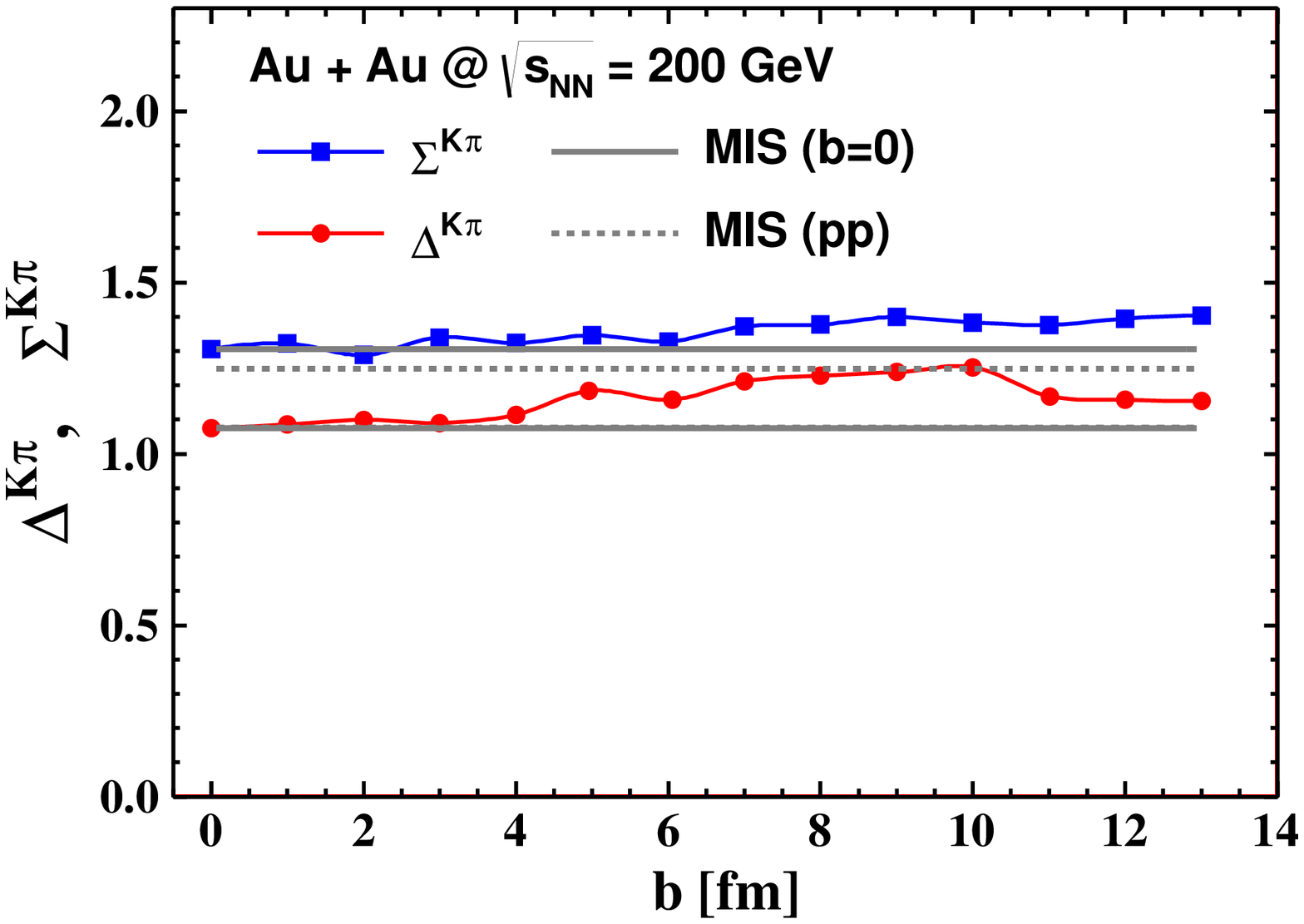,width=0.9\textwidth}
\caption{
The same as in Fig.~\ref{fig2} but for
$\sqrt{s_{NN}}=200$~GeV.}\label{fig4}
\end{figure}

The collision energy is chosen in the region where the NA49
Collaboration observes the horn structure~\cite{Horn}. The squares
and circles show the HSD results for pions and kaons,
respectively, at different impact parameters $b$. The stars
present the HSD results for $\omega_{part}$ (\ref{omegas}) which
correspond to the fluctuation of the number of nucleon
participants.
The lines are the results of the MIS. We consider two version of
MIS based on Eqs.~(\ref{n-pp},\ref{omega-pp}) and
Eqs.~(\ref{n-b0},\ref{omega-b0}), respectively, as discussed in
the previous section. The parameters $n_\pi$, $n_K$,
$\omega^*_\pi$, $\omega^*_K$, and $\rho_{K\pi}^*$ are thus taken
from the results of p+p collisions (\ref{n-pp},\ref{omega-pp}) and
from the HSD results at $b=0$ (\ref{n-b0},\ref{omega-b0}). These
MIS results will be denoted as MIS(pp) and MIS($b=0$),
respectively. The MIS($b=0$) calculations assume that
$\omega_{part}\cong 0$ at $b=0$. This is indeed supported by the
HSD results presented in Fig.~\ref{fig1}(b). The MIS parameters
(\ref{n-pp},\ref{omega-pp}) or (\ref{n-b0},\ref{omega-b0}) are
then used at all impact parameters $b$. The average number of
participants $\langle N_{part}\rangle$ and its fluctuations
$\omega_{part}$ are found for different values of $b\ge 0$ from
the HSD simulations.
Fig.~\ref{fig1}(a) demonstrates the ratios $\langle
\pi\rangle/\langle N_{part}\rangle$ and $\langle K\rangle/\langle
N_{part}\rangle$ as functions of $b$. One observes a slight
decrease of pion and kaon multiplicities per participating nucleon
with increasing $b$. On the other hand, the fluctuations of
$N_{part}$ strongly increase with $b$ as seen from
Fig.~\ref{fig1}(b). Thus according to the MIS formula
(\ref{omegaWNM}) one may expect an increase of $\omega_K(b)$ and
$\omega_\pi(b)$ as functions of $b$. This is indeed observed in
Fig.~\ref{fig1}(c) and Fig.~\ref{fig1}(d).
The MIS lines in Fig.~\ref{fig1}(c) and (d) give a correct
qualitative description of the HSD results for the pion and kaon
multiplicity fluctuations at different centralities in Au+Au
collisions (more details on the connection of the model of
independent sources and HSD results for the e-by-e fluctuations
can be found in Ref.~\cite{KGBG:2010}). Note that the results of
MIS(pp) and MIS($b=0$) are rather close to each other. As the
parameters (\ref{n-b0},\ref{omega-b0}) are fixed according to the
HSD  at $b=0$,  the MIS($b=0$) lines coincide, by construction,
with the HSD results at $b=0$. The $b$-dependence of the MIS
results is fully defined by the $b$-dependence of $\langle
N_{part}\rangle$ and $\omega_{part}$. Consequently, one may
conclude that a strong rise in the scaled variances $\omega_\pi$
and $\omega_K$ with $b$ seen in Fig.~\ref{fig1}(c) and
Fig.~\ref{fig1}(d) is caused by an increase of the e-by-e
fluctuations of $N_{part}$ with increasing $b$.

The participant number fluctuations are, however, cancelled out in
the strongly intensive measures. Therefore, $\Delta^{K\pi}$
(\ref{Delta}) and $\Sigma^{K\pi}$ (\ref{Sigma}) demonstrate only a
very weak $b$-dependence as seen in Fig.~\ref{fig2}. Note that the
MIS results (\ref{Delta1},\ref{Sigma1}) do not depend on $\langle
N_{part}\rangle$ and on $\omega_{part}$. Therefore, the MIS
results correspond to the values of $\Sigma$ and $\Delta$ which
are independent of impact parameter $b$. These MIS results are
presented by the horizontal lines in Fig.~\ref{fig2}.

The results presented in Figs.~\ref{fig1} and \ref{fig2} remain
qualitatively the same at higher collision energies. In
Fig.~\ref{fig3} and \ref{fig4} we present the corresponding
results in Au+Au collisions at the highest RHIC energy
$\sqrt{s_{NN}}=200$~GeV.
The kaon $\langle K\rangle/\langle N_{part}\rangle$ and pion
$\langle \pi \rangle/\langle N_{part}\rangle$ multiplicities per
participating nucleon increase with collision energy. As seen from
Fig.~\ref{fig3}(a), these values at $\sqrt{s_{NN}}=200$~GeV are
approximately 5 and 10 times larger than the corresponding values
at $\sqrt{s_{NN}}=7.7$~GeV presented in Fig.~\ref{fig1}(a). The
MIS thus predicts a much stronger increase of $\omega_K$ and
$\omega_\pi$ with $b$ at high collision energy. The HSD results
presented in Figs.~\ref{fig3}(c) and \ref{fig3}(d) support these
expectations. One observes also an increase of $\omega_K^*$ and
$\omega^*_\pi$ with $\sqrt{s_{NN}}$ reported earlier in
Ref.~\cite{KGB:2007}.

The results presented in Fig.~\ref{fig1} and Fig.~\ref{fig3} lead
to the conclusion that the fluctuation measures $\omega_\pi$ and
$\omega_K$ are useless for the samples of A+A collision events
within  wide centrality windows because of  large fluctuations of
the number of participants. For example, in a sample of  minimum
bias A+A collision events one would obtain very large values of
the scaled variances $\omega_K$ and $\omega_{\pi}$.
However, the dominant contributions to these values of the scaled
variances come evidently from the participant number fluctuations.
Indeed, the HSD results correspond to a huge value of
$\omega_{part}\approx 100$
in a sample of the minimum bias Au+Au (or Pb+Pb) collisions.
Thus, for $i$th hadron species (e.g., $i=\pi,K$) very large
contributions $n_i\omega_{part}$  come to $\omega_i$. These
contributions increase with collision energies due to an increase
of $n_i$.
On the other hand, the contributions of participant number
fluctuations are approximately cancelled out in (\ref{Delta}) and
(\ref{Sigma}), and the strongly intensive measures $\Delta$ and
$\Sigma$ are expected to remain close to their numerical
values at $b=0$. This expectation is supported by the
HSD results presented in Figs.~\ref{fig2} and \ref{fig4}.

\begin{figure}[ht!]
 \epsfig{file=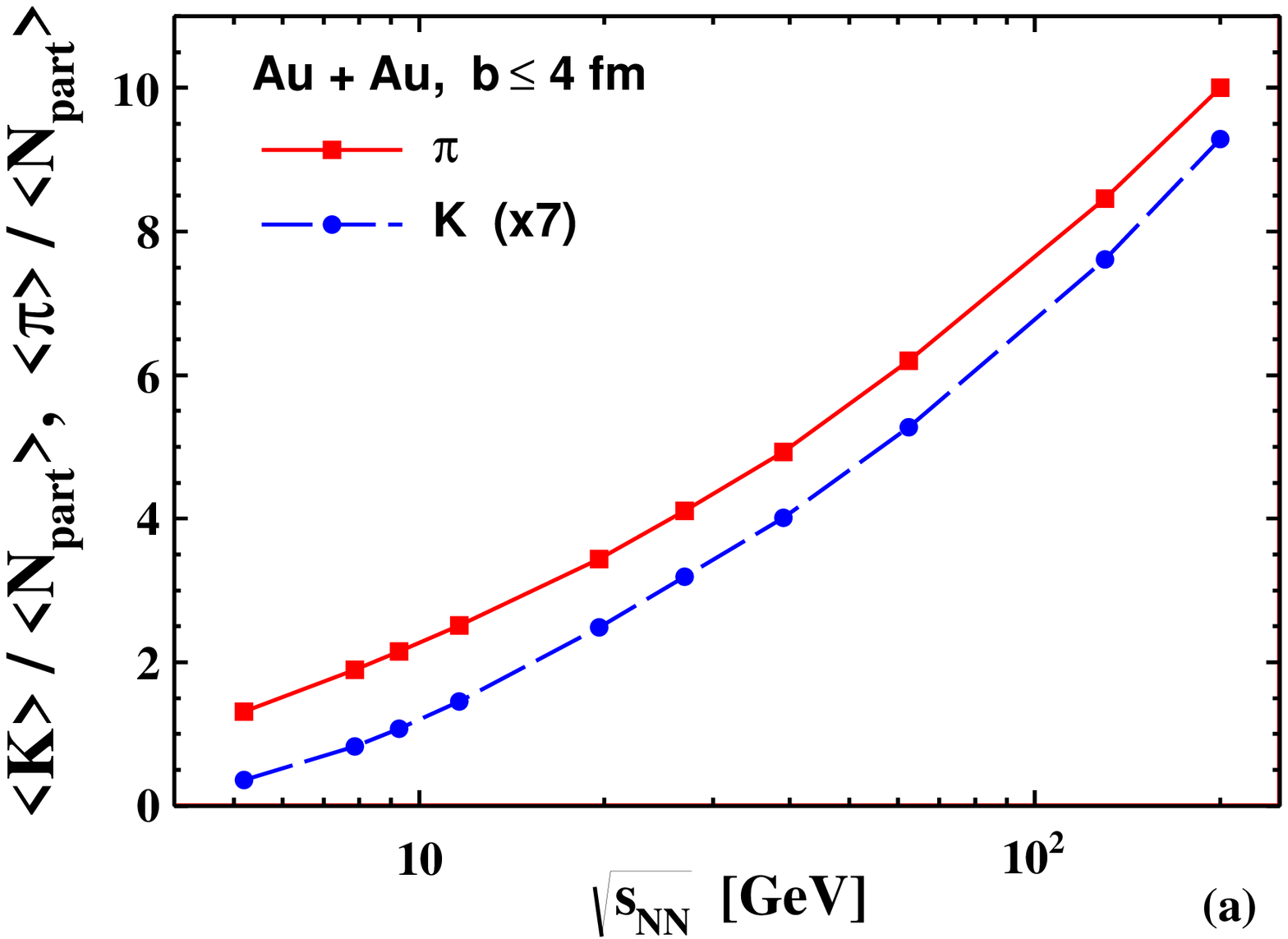,width=0.45\textwidth}
 \epsfig{file=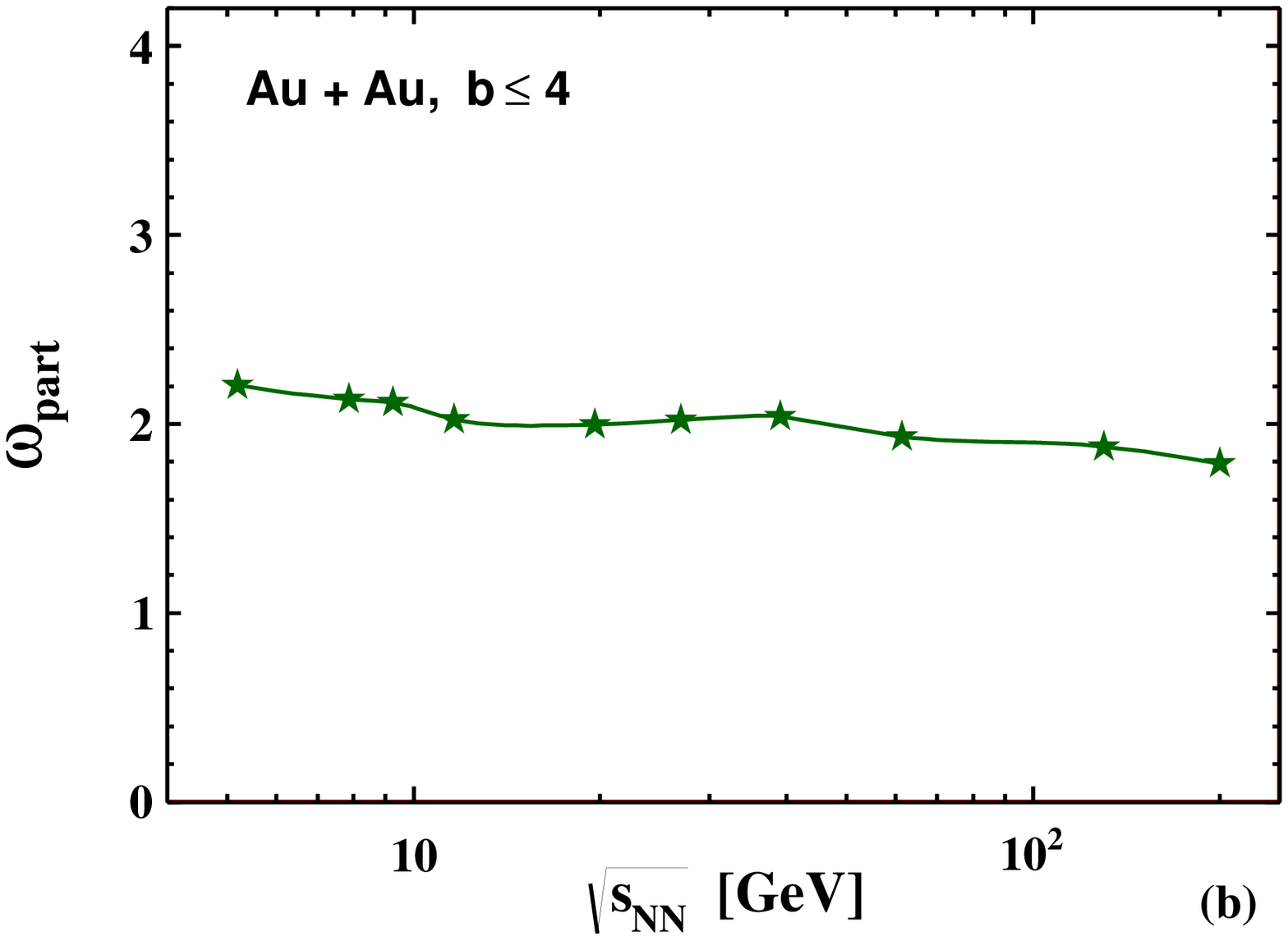,width=0.45\textwidth}
\caption{
The HSD results in Au+Au collisions for
$b\le 4$~fm as a function of the center of mass energy per nucleon
pair $\sqrt{s_{NN}}$. (a): The values of relative particle
multiplicities per participating nucleon  $\langle
K\rangle/\langle N_{part}\rangle$  and $\langle \pi\rangle/\langle
N_{part}\rangle$. Note that $\langle K\rangle/\langle
N_{part}\rangle$ is multiplied by a factor of 7. (b): The scaled
variances  $\omega_{part}$.
 }
 \label{fig5}
\end{figure}

\begin{figure}[ht!]
\epsfig{file=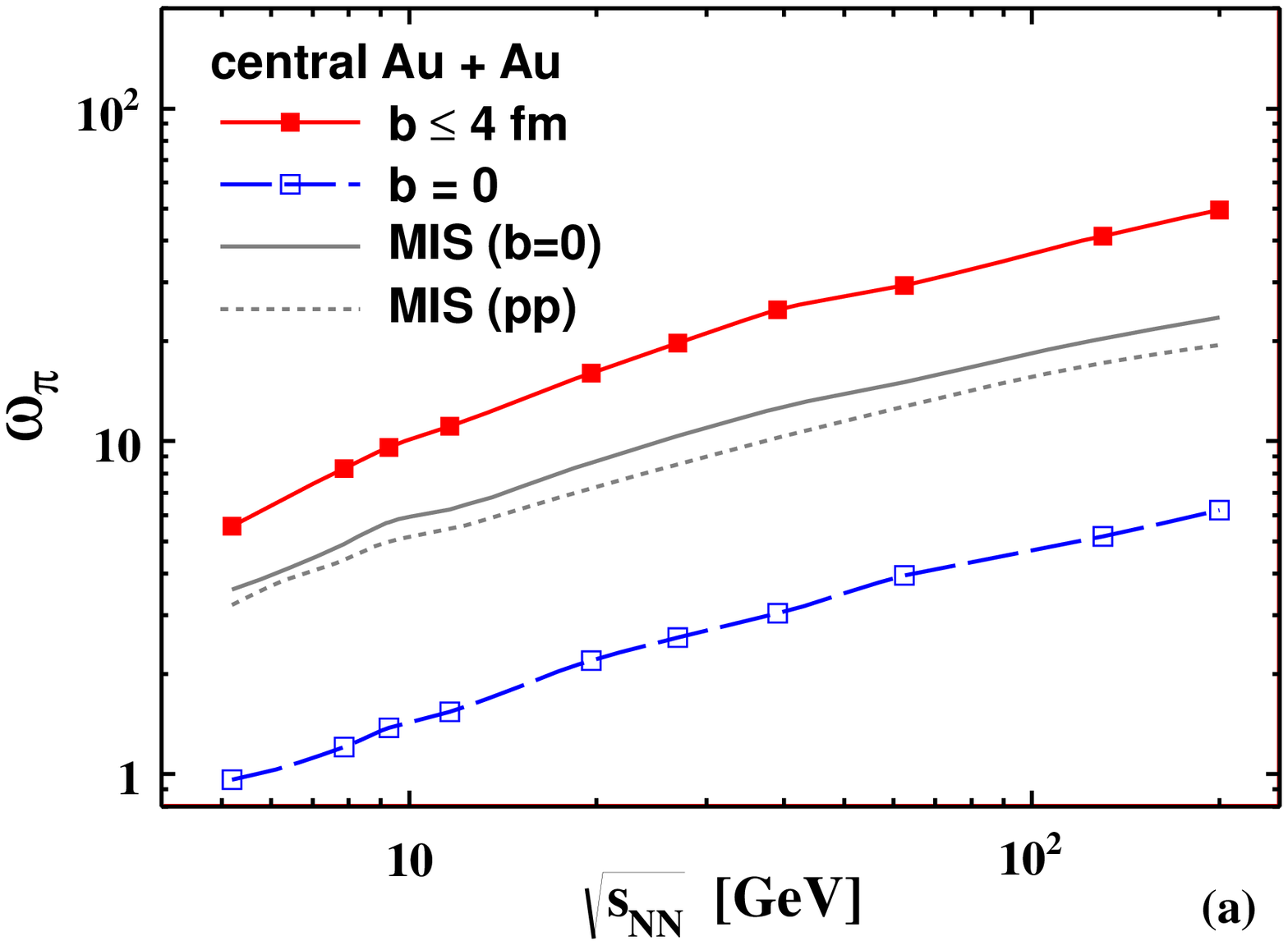,width=0.45\textwidth}
\epsfig{file=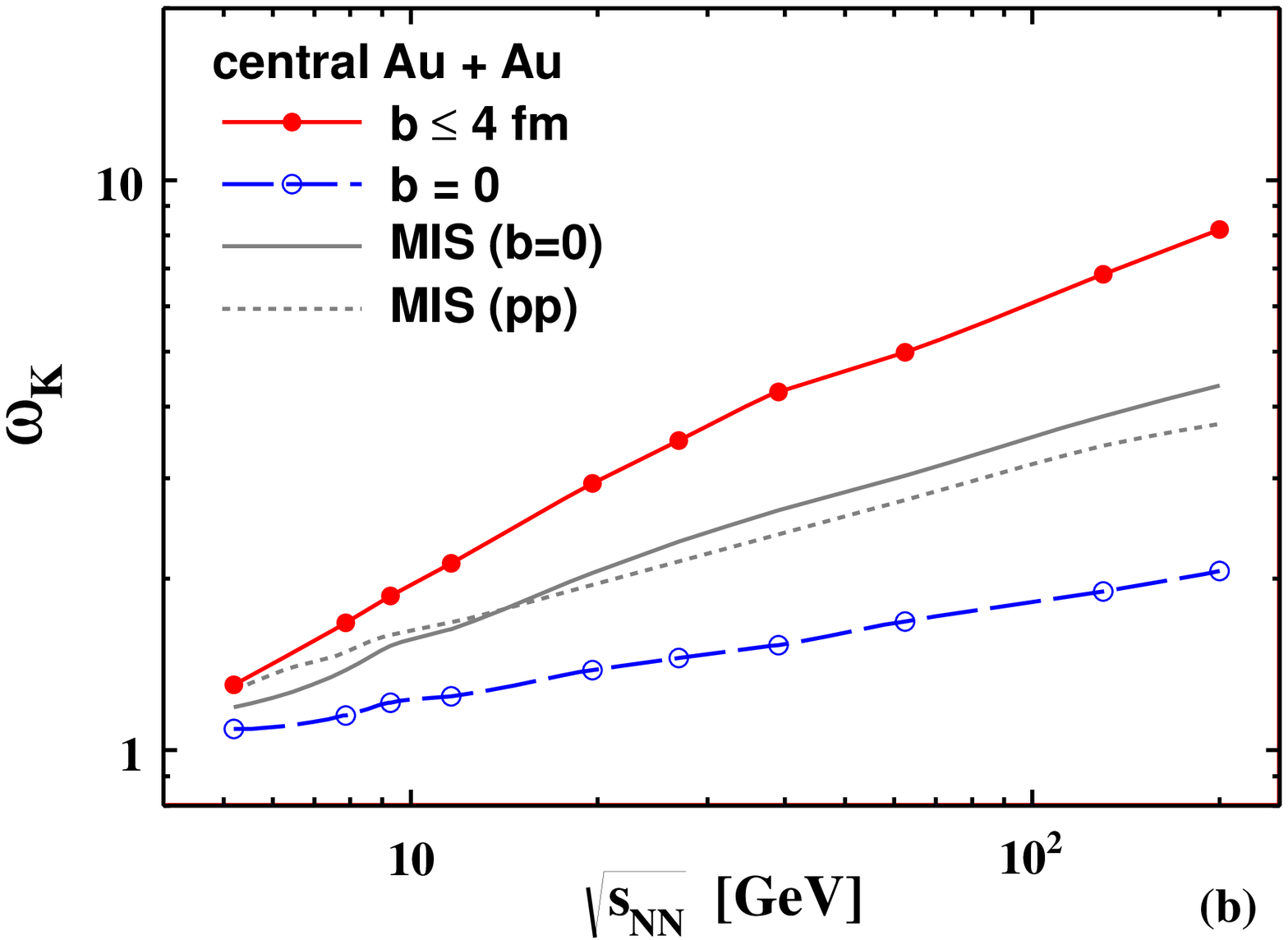,width=0.45\textwidth}
\caption{
The HSD results for the scaled variances
$\omega_\pi$ (a) and $\omega_K$ (b) in Au+Au collisions at $b=0$
(open symbols) and at $b\leq 4$~fm (full symbols).
The solid and dotted lines show the MIS($b=0$) and MIS(pp)
results, respectively.
\label{fig6} }
\end{figure}

\begin{figure}[ht!]
\epsfig{file=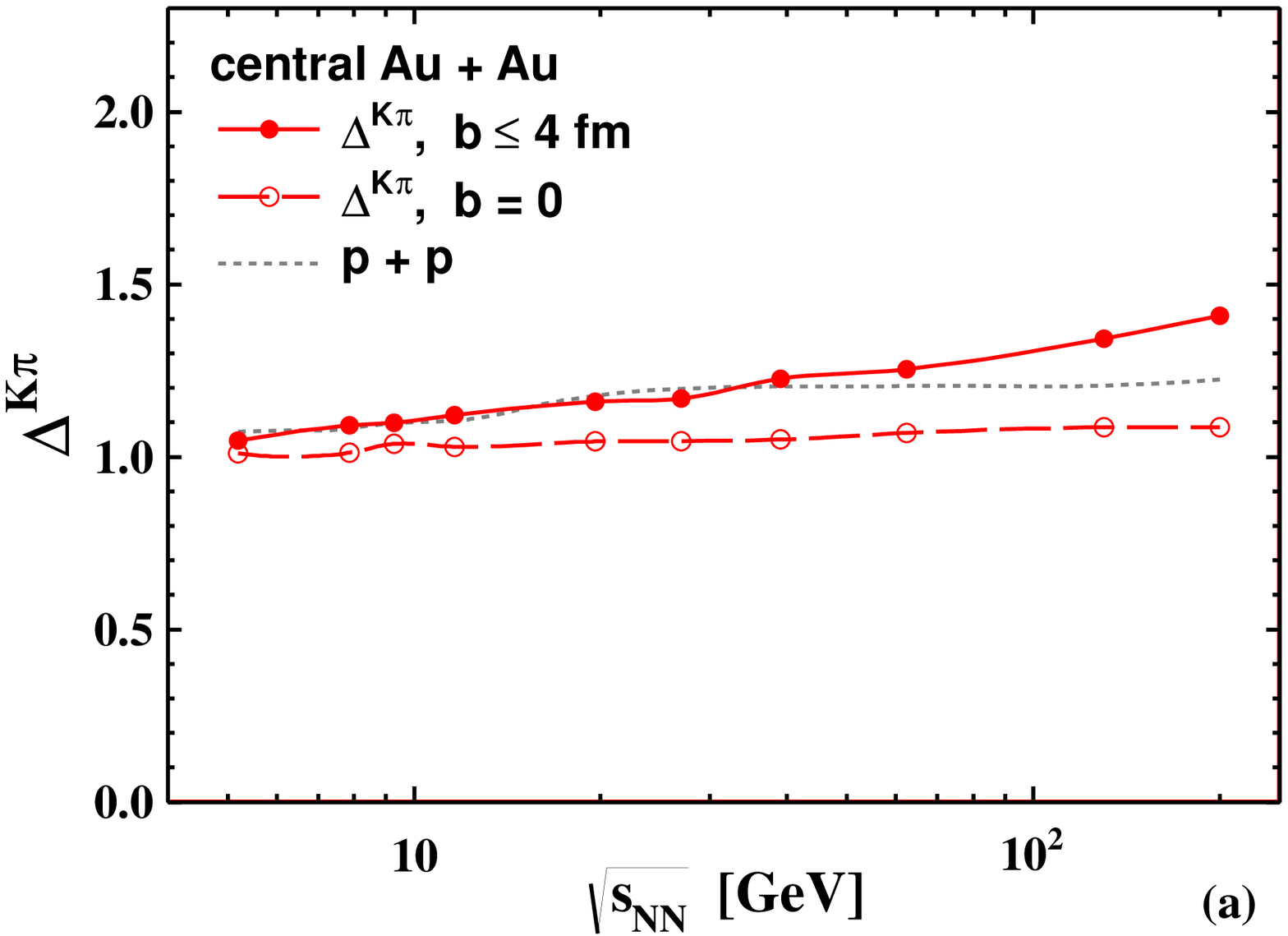,width=0.49\textwidth}
\epsfig{file=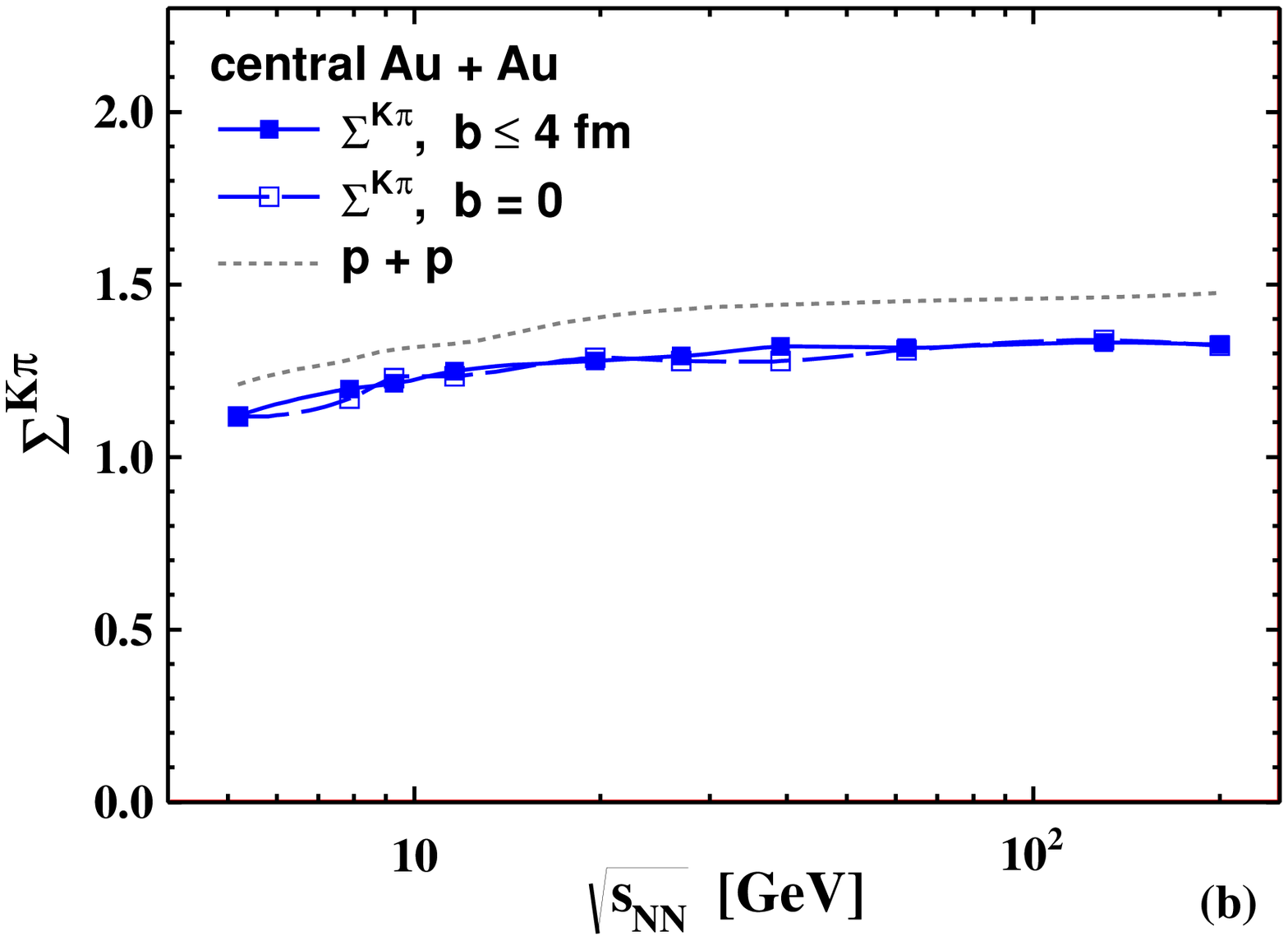,width=0.49\textwidth}
 \caption{
 The HSD results
for  the strongly intensive measures $\Delta^{K\pi}$ and
$\Sigma^{K\pi}$ in Au+Au collisions at $b=0$ (open symbols) and at
$b\leq 4$~fm (full symbols). The lower and upper dotted lines show
the MIS(pp) results for $\Delta^{K\pi}$ and $\Sigma^{K\pi}$,
respectively. \label{fig7} }
\end{figure}

As a further test we perform the HSD simulations in Au+Au
collisions at different collision energies for $b\le 4$~fm . This
requirement corresponds approximately to 10\% of the most central
collisions.
%
For the sample of collision events with $b\le 4$~fm, the
dependence on collision energy of relative multiplicities $\langle
\pi\rangle/\langle N_{part}\rangle$ and $\langle K\rangle\langle
N_{part}\rangle$, and scaled variance $\omega_{part}$ is shown in
Fig.~\ref{fig5}(a) and \ref{fig5}(b), respectively.
In Fig.~\ref{fig6} the HSD results are presented for the scaled
variances of $\omega_K$ and $\omega_\pi$ as a function of the
collision energy $\sqrt{s_{NN}}$.
The calculations in Au+Au collisions for $b\leq 4$~fm are compared
to those for $b=0$ and to the results of MIS.
The main features of the results for the scaled variances
$\omega_\pi$ and $\omega_K$ presented in Fig.~\ref{fig6} can be
summarized as follows: An averaging over Au+Au collision events
with $b\le 4$~fm leads to  very strong increase of $\omega_\pi$
and $\omega_K$ in a comparison to their values at $b=0$. For
example, at $\sqrt{s_{NN}}=200$~GeV the scaled variance
$\omega_\pi$ for $b\le 4$~fm is higher than $\omega_\pi$ for $b=0$
by approximately a factor of 10.  As seen from Fig.~\ref{fig6},
the MIS results explain only a part of this increase.

The contributions from participant number fluctuations  are quite
strong in the sample of Au+Au collision events with $b\le 4$~fm.
These contributions are, however, cancelled out in the strongly
intensive measures $\Delta^{K\pi}$ (\ref{Delta}) and
$\Sigma^{K\pi}$ (\ref{Sigma}). These measures for the sample $b\le
4$~fm remain close to their numerical values at $b=0$. This is
shown in Fig.~\ref{fig7}. Note that the MIS($b=0$) results for
$\Delta^{K\pi}$ and $\Sigma^{K\pi}$ in the sample of $b\le 4$~fm
are identical to those at $b=0$. Therefore, only the p+p HSD
results are presented in Fig.~\ref{fig7} by the dotted lines.
Figure~\ref{fig7}(b) shows that $\Sigma^{K\pi}(b\le
4$~fm)~$\cong~\Sigma^{K\pi}(b=0)$. On the other hand, an increase
of about 30\% at the highest RHIC energy $\sqrt{s_{NN}}=200$~GeV
is seen in Fig.~\ref{fig7}(a) for $\Delta^{K\pi}$ for the sample
of $b\le 4$~fm in comparison to that at $b=0$ .

\section{Comparison to the NA49 and STAR data}\label{sec-Exp}

In this section we present a comparison of the HSD results for the
strongly intensive measures $\Delta^{K\pi}$ and $\Sigma^{K\pi}$
with existing data on e-by-e fluctuations.

\begin{figure}[ht!]
 \epsfig{file=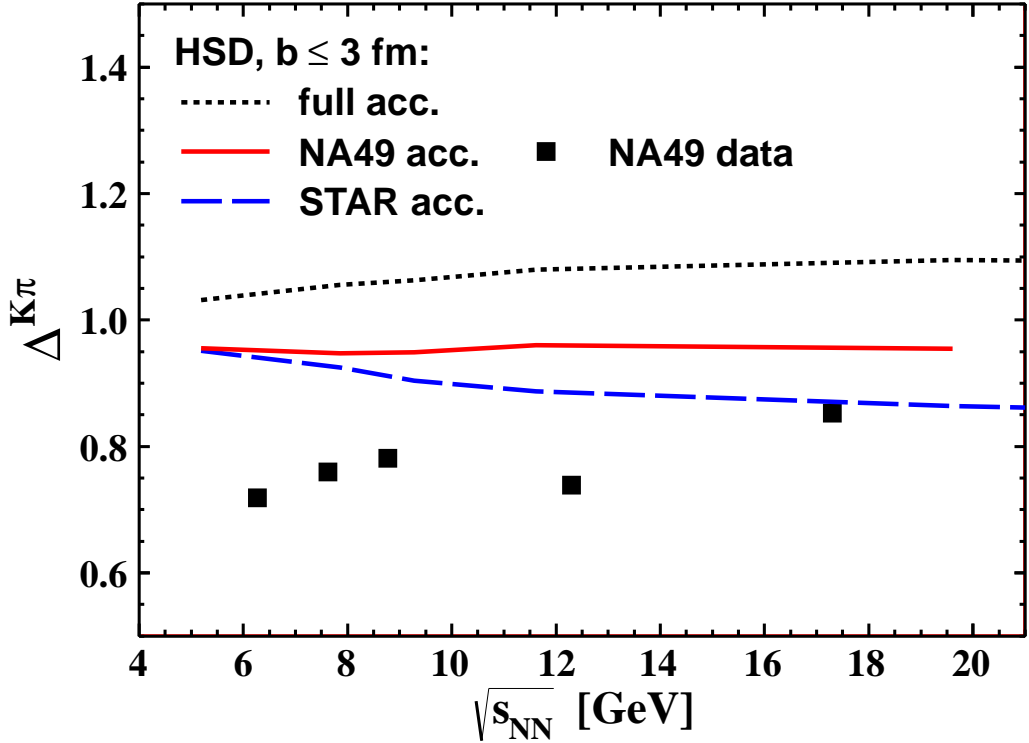,width=0.9\textwidth}
 \caption{
The squares are the NA49 data for $\Delta^{K\pi}$ for 3.5\% most
central Pb+Pb collisions. The solid, dashed, and dotted lines show
the HSD results in Au+Au collisions at $b\leq 3$fm within the
NA49, STAR, and full $4\pi$ acceptances, respectively.
} \label{fig-Delta}
\end{figure}
\begin{figure}[ht!]
 \epsfig{file=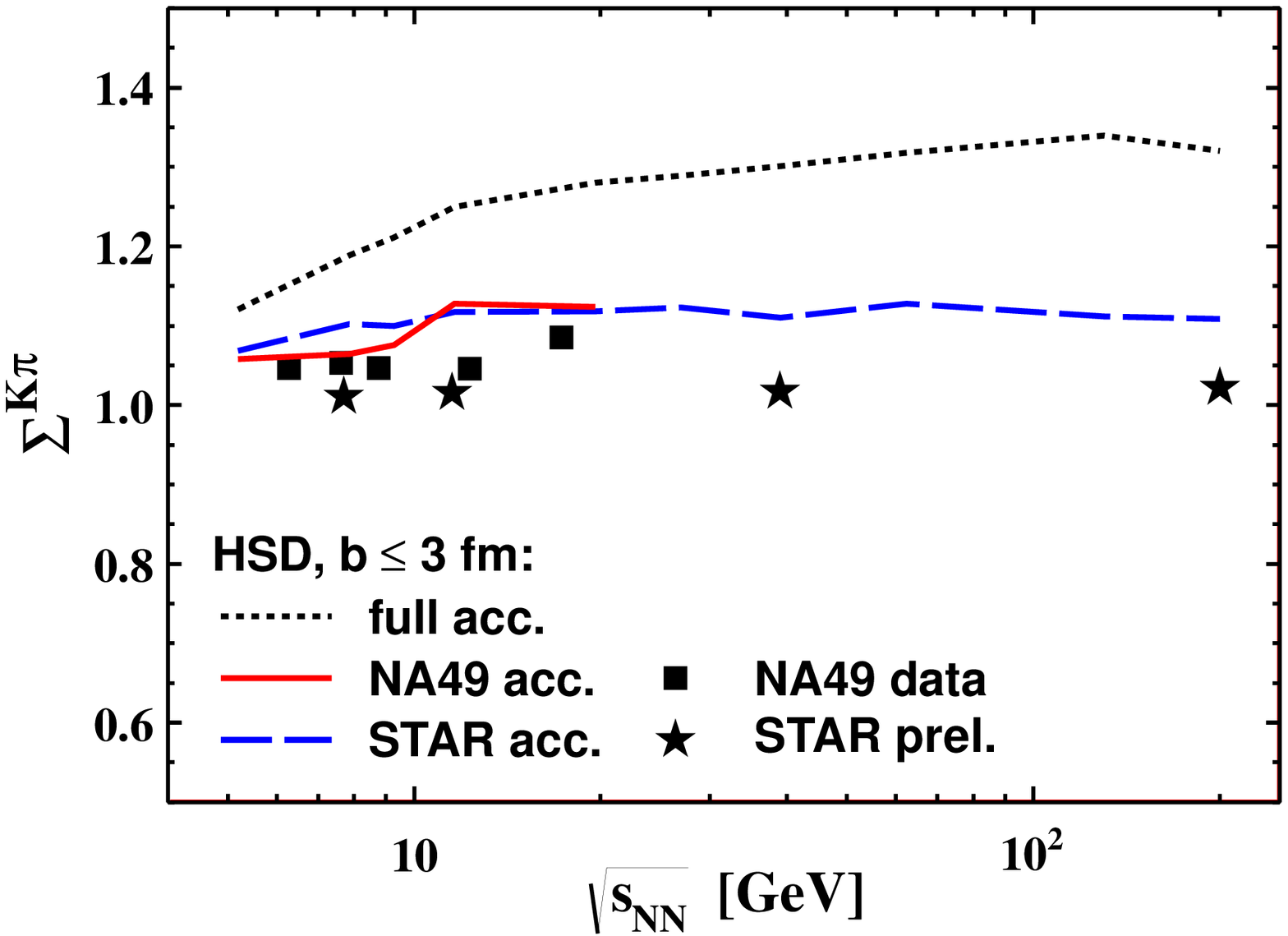,width=0.9\textwidth}
 \caption{
The squares are the NA49 data for $\Sigma^{K\pi}$ for 3.5\% most
central Pb+Pb collisions, and the stars are the STAR data for 5\%
most central Au+Au collisions. The solid, dashed, and dotted lines
show the HSD results in Au+Au collisions at $b\leq 3$fm within the
NA49, STAR, and full $4\pi$ acceptances, respectively.
} \label{fig-Sigma}
\end{figure}

Recently the NA49 Collaboration published the data on mean
multiplicities, correlations and fluctuations of pions, kaons and
protons in Pb+Pb collisions~\cite{Rustamov}. From these data we
are able to construct the fluctuation measure $\Delta^{K\pi}$ and
$\Sigma^{K\pi}$. The STAR Collaboration has published only the
data for $\nu_{dyn}$~\cite{Tarnowsky} in Au+Au collisions. We
manage to recalculate them into $\Sigma$ values using
Eq.~(\ref{nu-Psi}) and the preliminary data on mean multiplicities
of kaons and pions~\cite{Tarnowsky2}.

The results for $\Delta^{K\pi}$ in the SPS energy region are
presented in Fig.~\ref{fig-Delta}. The  squares present the NA49
results for 3.5\% most central Pb+Pb collisions.  One can see that
the NA49 data for $\Delta^{K\pi}$ show a non-monotonous behavior
with a bump at $\sqrt{s_{NN}}=8.8$~GeV and a deep at
$\sqrt{s_{NN}}=12.3$~GeV.  The NA49 Collaboration did not yet
publish the error-bars for the first and  second moments of $K$
and $\pi$ multiplicity distributions. Therefore, it is difficult
to conclude whether the non-monotonous structure in
$\Delta^{K\pi}$ has a real statistical significance. However, this
potential irregularity happens in the energy range where other
signals of unusual behavior ({\it kink, horn, step}) were observed
by the NA49 Collaboration~\cite{Horn}. It would be interesting to
see the STAR data for $\Delta^{K\pi}$ in this energy range.

The results for $\Sigma^{K\pi}$ are presented in
Fig.~\ref{fig-Sigma}. The squares correspond to the NA49 data and
the stars to the STAR data. The available error-bars for
$\nu_{dyn}^{K\pi}$ can be used for a rough estimate of the
error-bars for $\Sigma^{K\pi}$. These error-bars seem to be quite
small and comparable with the size of the star-symbols in
Fig.~\ref{fig-Sigma}.

The NA49 and STAR Collaborations have different colliding nuclei
(Pb+Pb and Au+Au, respectively), different acceptances and
different centralities (3.5\% and 5\% most central collisions,
respectively). We restrict our HSD simulations to impact
parameters $b\leq 3$ in Au+Au collisions. This corresponds
approximately to 5\% most central events. We use this centrality
criterion for both NA49 and STAR data, as the quantities
$\Delta^{K\pi}$ and $\Sigma^{K\pi}$ are only weakly dependent on
centrality selection. However, we take into account the exact
experimental acceptances which are different for the NA49 and STAR
data. In Figs.~\ref{fig-Delta} and \ref{fig-Sigma} the HSD results
are shown by the solid and dashed lines for the NA49 and STAR
acceptances, respectively. The HSD results in the full
$4\pi$-acceptance are shown by the dotted lines. A presence of the
dashed lines in Figs.~\ref{fig-Delta} and \ref{fig-Sigma} helps to
estimate the effects of the limited acceptances in NA49 and STAR
experiments.

As seen in Figs.~\ref{fig-Delta} and \ref{fig-Sigma} the HSD
results correspond to higher values of $\Delta^{K\pi}$ and
$\Sigma^{K\pi}$ than their experimental estimates. This is
especially seen for $\Delta^{K\pi}$. The HSD results depend
monotonously on collision energy and can not explain the
bump(deep) in the NA49 data for $\Delta^{K\pi}$. Therefore, the
origin of this `bump' (if it will survive in future measurements)
is not connected with simple geometrical or limited acceptance
effects, as these effects are taken into account in the HSD
simulations.

\section{Summary}\label{sec-sum}
The recently proposed two families of strongly intensive measures
of fluctuations and correlations are studied within the HSD
transport approach to nucleus-nucleus collisions. We test the
measures $\Delta^{K\pi}$ (\ref{Delta}) and $\Sigma^{K\pi}$
(\ref{Sigma}) for the fluctuations of kaon and pion numbers in
Au+Au collisions at different collision energies and different
centralities.

The conventional measures like scaled variances $\omega_K$ and
$\omega_{\pi}$ become useless for wide centrality samples because
of the dominant contribution from the participant number
fluctuations. This fact required a very rigid centrality selection
like 1\% most central Pb+Pb collision events in Ref.~\cite{Lung}.
The other popular measure, $\nu_{dyn}$, is independent of
participant number fluctuations, but depends on the average number
of participants. Therefore, $\nu_{dyn}$ is inconvenient for
comparison of p+p and Au+Au collisions as well as for the search
for the QCD critical point by system size scan program of NA61
Collaboration at the CERN SPS.

The quantities $\Delta^{K\pi}$ and $\Sigma^{K\pi}$ appear to be
useful measures of chemical fluctuations in the wide centrality
samples of collision events. In the sample of 10\% most central
Au+Au collision events we find that $\Delta^{K\pi}$ is slightly
larger than that at $b=0$, and $\Sigma^{K\pi}$ is approximately
equal to its value at zero impact parameter $b=0$.
This makes $\Delta^{K\pi}$ and $\Sigma^{K\pi}$ rather helpful in
studies of event-by-event fluctuations in nucleus-nucleus
collisions. Combining existing experimental data of NA49 and STAR
Collaborations on $K$-$\pi$ fluctuations and correlations we have
obtained $\Delta^{K\pi}$ and $\Sigma^{K\pi}$ and compared them to
the corresponding HSD calculations. The data on $\Sigma^{K\pi}$
depend monotonously on the collision energy, whereas
$\Delta^{K\pi}$ from the NA49 data has a little bump(deep) in the
region $\sqrt{s_{NN}}\cong 8\div 12$~GeV, where other signals of
irregular behavior of physical quantities were previously
reported~\cite{Horn}.
The HSD describes $\Sigma^{K\pi}$ reasonably well,  but does not
reproduce the behavior of $\Delta^{K\pi}$. The data analyzed in
the present paper correspond to $3\div 5$\% most cental
collisions. This centrality window can be enlarged at least to
10\% in the future experimental studies using strongly intensive
measures $\Delta^{K\pi}$ and $\Sigma^{K\pi}$.

\begin{acknowledgments} We are thankful to W.~Cassing,
M.~Ga\'zdzicki, K.~Grebieszkow, W.~Greiner,  M. Ma\'ckowiak,
A.~Rustamov, and  T.~Tarnowsky,  for fruitful discussions and
comments. E.L.B. and V.V.B. acknowledge the financial support
through the "HIC for FAIR" framework of the "LOEWE" program.
V.V.B. and V.P.K. acknowledge also the financial support from the
DFG foundation. The work of M.I.G. was supported by the Humboldt
Foundation and by the Program of Fundamental Research of the
Department of Physics and Astronomy of NAS, Ukraine.
\end{acknowledgments}

\end{document}